\documentclass[aps,twocolumn,amsmath,amssymb,prl]{revtex4-1}
\usepackage{amssymb}
\usepackage{graphicx}
\usepackage{epsfig}
\usepackage[latin1]{inputenc}
\usepackage{dcolumn}
\usepackage{bm}
\begin{document}

\title{Quantum fluctuations, the Boson peak, and the glassy state of biomolecules}
\author{T. A.  Lima}
\author{M. S. Ishikawa}
\author{H. S. Martinho}
\email{herculano.martinho@ufabc.edu.br}

\affiliation{Centro de Ciências Naturais e Humanas, Universidade Federal do ABC, Av. dos Estados 5001, Santo André-SP, 09210-580, Brazil}

\begin{abstract}
 It has been recognized in the literature that some physical properties of hydrated biomolecules, e.g., the occurrence of Boson peak , resembles of those of glassy state.  In the present work is shown that quantum fluctuations play a fundamental role on describing the glassy state of biomolecules, specially at lower hydration levels. It is reported a remarkable linear dependence on the quantumness and the slope of the Boson peak frequency temperature dependence which would be used to classify de degree of quantum contributions to the glassy state by glasses in general. Finally, it is shown that the Boson peak two-bands spectral structure observed in some cases could be direct linked to the anisotropy of the material elastic properties.
\end{abstract}

\maketitle

It has been reported some intriguing similarities between biological macromolecules and glasses.  One could cite the observation of an anomalous low temperature specific heat, the characteristic slow energy relaxation processes, the occurrence of dynamical transitions, and the excess in the low frequency phonon density of states observed in the dynamical spectra by inelastic neutron and Raman scattering experiments (the so-called Boson Peak), as the main experimentally observed parallel findings between these two classes of materials.\cite{green1994protein,iben1989glassy,piazza2005slow,xie2001excited,shintani2008universal,khodadadi2010broad}

The Boson Peak ($BP$) is a distinctive feature of many glassy and disordered crystalline solids which was been also observed in macro-molecules like DNA and proteins.\cite{green1994protein,iben1989glassy,piazza2005slow,xie2001excited,shintani2008universal,khodadadi2010broad} But, recently it has been suggested that it may be correlated to anharmonic and glass transitions observed at $T_{D}\sim 180-200$ K and $T_{g}\sim 80-100$ K, respectively. As pointed out by Khodadadi \textit{et al} \cite{khodadadi2008influence} both have different physical origin and should not be confused.

Based on quasi-elastic neutron scattering (QENS) measurements on lysozyme, Chen \textit{et al.}\cite{chen2006observation} interpreted the transition at $T_{D}$ as a fragile-to-strong dynamic crossover where the structured water makes a transition from a high-density to a low-density state. This scenario was based on studies that advocated a second critical point of water (liquid-liquid critical point) in confined super-cooled water.\cite{xu2005relation} However, a recent high resolution QENS experiment performed by Doster \textit{et al.}\cite{doster2010dynamical} in fully deuterated C-phycocyanin protein, showed no evidence of such fragile-to-strong character at $T_{D}$. As pointed out by these authors, their experimental findings are consistent with a glassy transition of the hydration shell. However, the exact nature of the glassy state remains unclear. Moreover, the particular role of protein and water is still an issue.

Inelastic light (Raman) and neutron scattering techniques have been applied to probe the temperature and hydration dependence of the $BP$ of several glasses and biological macromolecules.\cite{shintani2008universal,khodadadi2010broad,leyser1999far,malinovsky1986nature,mantisi2010non,steffen1994depolarized} In general, the observed spectra  exhibited a broad central line with a weak shoulder between $15$ and $30$ cm$^{-1}$.\cite{leyser1999far,malinovsky1986nature,sokolov1994dynamics} Similar trend was also observed in specific heat data.\cite{mantisi2010non,surovtsev2001evaluation} The $C/T^{3}$ against $T$ plot presented a low temperature ($T\sim 15$ K) anomaly related to the $BP$.\cite{mantisi2010non,surovtsev2001evaluation}

Studies on short protein fragments such as amino acids could provide a very interesting approach to solve the problem. The low molecular weighting amino acids are very suitable for computational modeling due to the comparatively small structures and can be used as functional prototypes of more complex structures such as proteins and lipids. In fact, subtle fluctuations in their physical properties could trigger the most significant changes detected in large macromolecules.\cite{giuffrida2006role}

Accordingly, the present paper is devoted to investigate the nature of the possible glassy state in biomolecules probing the low-frequency ( $\omega<50$ cm$^{-1}$) Raman scattering in hydrated and almost dry L-Cysteine amino acids.

The samples consist in amino acid  L-Cysteine almost dry ($0.035$) (Sigma-Aldrich, purity $> 99\%$) and hydrated, to $0.57$ and$0.73$ g water/g amino acid which will be labeled as LCys$:0.035$, LCys$:0.57$, and LCys$:0.73$. The Raman measurements were carried out using a triple spectrometer (T64000, HORIBA Jobin-Yvon) with a thermoelectric cooled CCD detector (Synapse, HORIBA Jobin-Yvon). The 532 nm line of an optically pumped semiconductor laser (Verdi G5, Coherent) was used as excitation source. The laser power at the sample was kept below 15 mW on a spot diameter of about 100  m. The samples were cooled in a cold finger of a 4K ultra-low-vibration He closed-cycle cryostat (CS-204SF-DMX-20-OM, Advanced Research System), and measured in a near-backscattering configuration. The calorimetric measurements were performed using the commercial automated heat-capacity measurement system option of the Physical Properties Measurements System (PPMS, Quantum Design). They were performed using the relaxation method.\cite{lashley2003critical}

\begin{figure}[tbh!]
\centering
  \includegraphics[height=4.5cm]{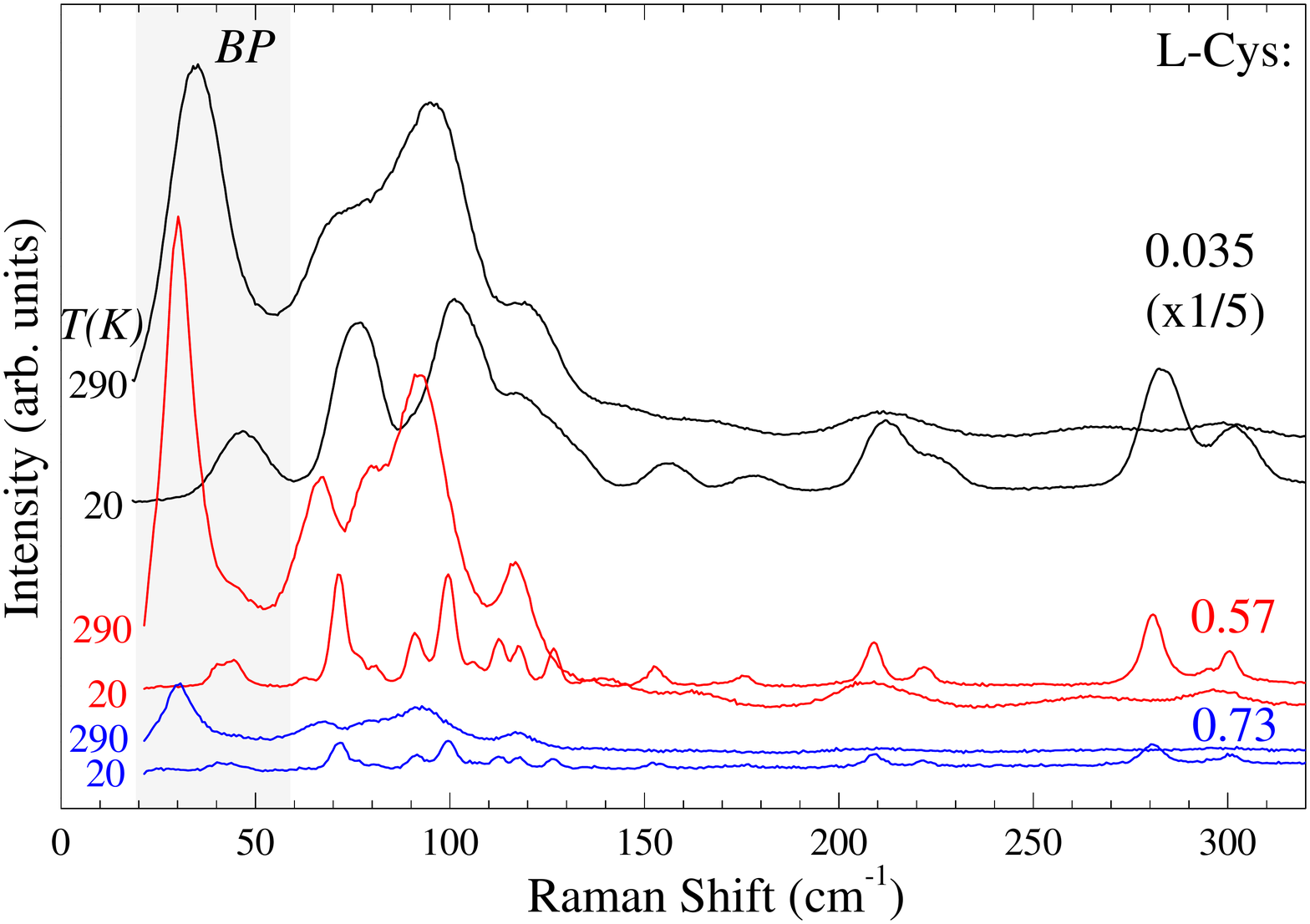}
  \caption{Raman spectra of LCys$:0.035,0.57$, and $0.73$ at $T=20$ K and $290$ K.  The spectral region related to the $BP$ at $20-55$ cm$^{-1}$ is also indicated.}
  \label{fig:1}
\end{figure}

Figure \ref{fig:1} shows the Raman spectra of LCys$:0.035$, LCys$:0.57$, and LCys$:0.73$ at $20$K and $290$ K. Following the vibrational modes assignment present on our previous work\cite{lima2012anharmonic} it was is possible to conclude that the band in the spectral region $20-60$ cm$^{-1}$ is not a Raman-active normal mode of the structure. This band narrowed, softened, and increased the intensity with temperature following the temperature dependence of the boson occupation factor  $n(\omega,T)=(exp(\hbar\omega/k_{B}T)-1)^{-1}$. The $BP$ presents similar thermal properties, as described in the literature.\cite{khodadadi2008influence,schirmacher2013boson} Thus one could assign the low frequency band in the $20-60$ cm$^{-1}$ spectral window to the $BP$.

\begin{figure}[tbh!]
\centering
  \includegraphics[height=4.5cm]{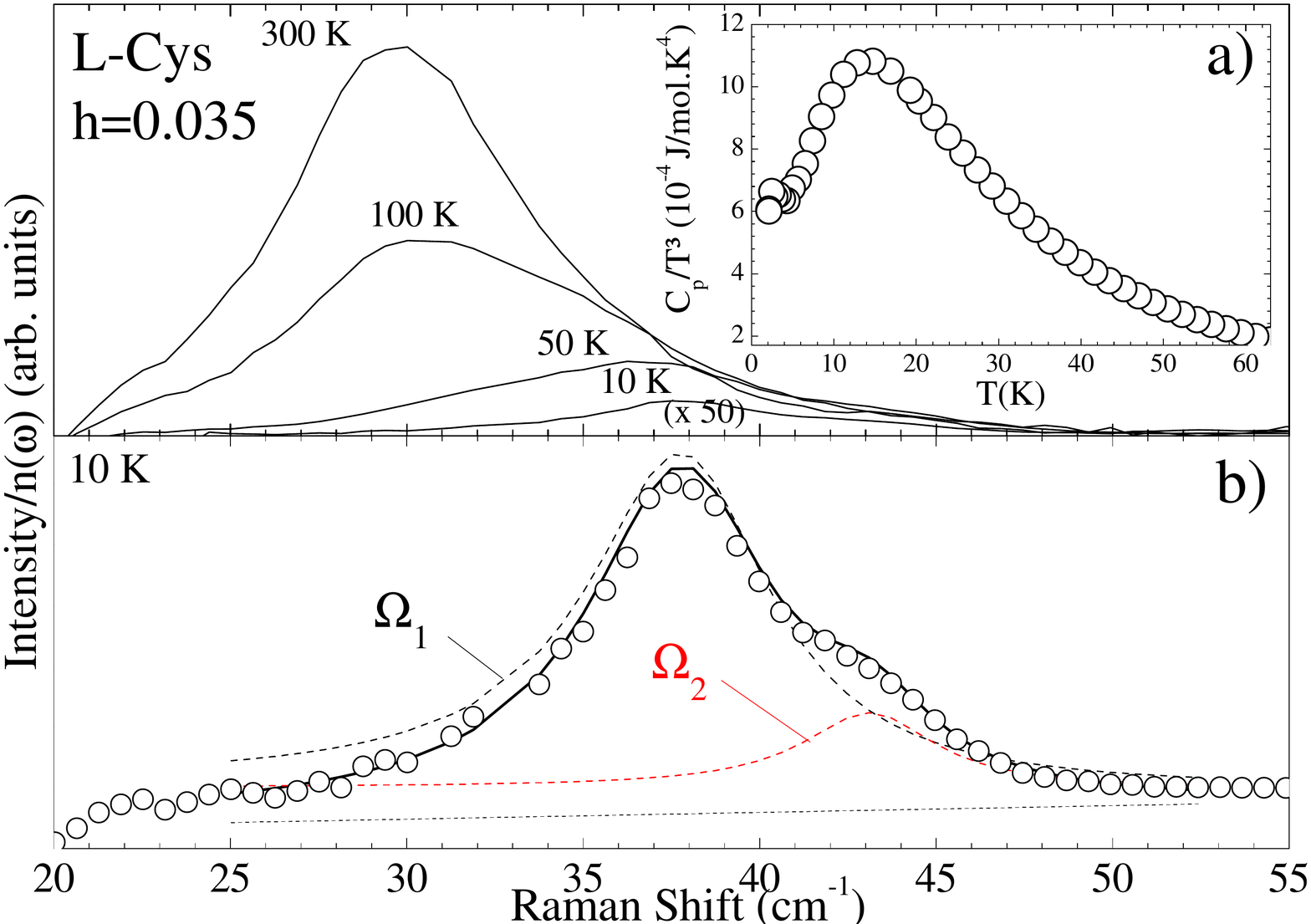}
  \caption{a) $BP$ at some selected temperatures ($10$K, $50$K, $100$K, and $300$ K) for L-Cys$: 0.035$). The inset shows the specific heat as $C_{p}/T^{3}$ versus $T$ plot showing near to $14$K related to the excess of VDOS. b) $BP$ at $T=10$ K data for L-Cys sample (open circles). The solid curve is a fit to eq. \ref{damped} where two contributions (dashed lines) at  $\Omega_{1}$ and  $\Omega_{2}$ frequencies where considered .}
  \label{fig:2}
\end{figure}

The temperature dependence of the $BP$ is better accomplished compensating normalizing  the spectra by the Bose-Einstein occupation factor. The data for $BP$ at some selected temperatures are shown on Fig. \ref{fig:2}a). The inset shows the anomaly in $C_{p}/T^{3}$  against $T$  data. The broad maximum observed at  $T\sim 15$ K is a manifestation of the emergence of the $BP$ in the vibrational density of states (VDOS).\cite{surovtsev2001evaluation}

Several models were proposed to explain both the $BP$ and $BP$ experimental measurements data on glasses. A review concerning it could be obtained in ref.\cite{Kob2011}. In particular, Leyser, Doster, and Dielr\cite{leyser1999far} concluded about the relevance of viscoelastic damping of modes in the $BP$ region in macromolecules as  globular protein. Spatial fluctuations of the shear modulus were  invoked to explain the anomalies related to $BP$ by Schirmacher \cite{schirmacher2013boson}. Moreover, the damped librational character of a side-chain oscillations at low frequencies of proteins had been well described in the literature.\cite{leyser1999far,diehl1997water,kneller1994liquid} Thus one could argue that a damped harmonic oscillator model is a suitable way to express the inelastic light scattering related to the $BP$,

\begin{equation}\label{damped}
    I(\omega)=A\frac{\Omega^{3}\gamma}{(\omega^{2}-\Omega^{2})^{2}+(\omega\Omega^{2}\gamma)^{2}}
\end{equation}
were $\Omega $ is the $BP$ frequency.  The parameter $\gamma$ could be interpreted as the Newtonian friction once considering eq. \ref{damped} as the $q=0$ elastic limit of the  Leyser, Doster, and Dielr model (eq. (1) of \cite{leyser1999far}). Khodadadi \textit{et al.}\cite{khodadadi2010broad} used a model similar to eq. \ref{damped} to fit their low frequency $BP$ Brillouim experimental data. Figure \ref{fig:2}b) shows the $10$ K spectrum of LCys$:0.035$ fitted to eq.\ref{damped}. Similar to findings of some previous works in macromolecules  the $BP$ presented as convolution of two bands at $\Omega_{1}$ and $\Omega_{2}$.

\begin{figure}[tbh!]
\centering
  \includegraphics[height=5.5cm]{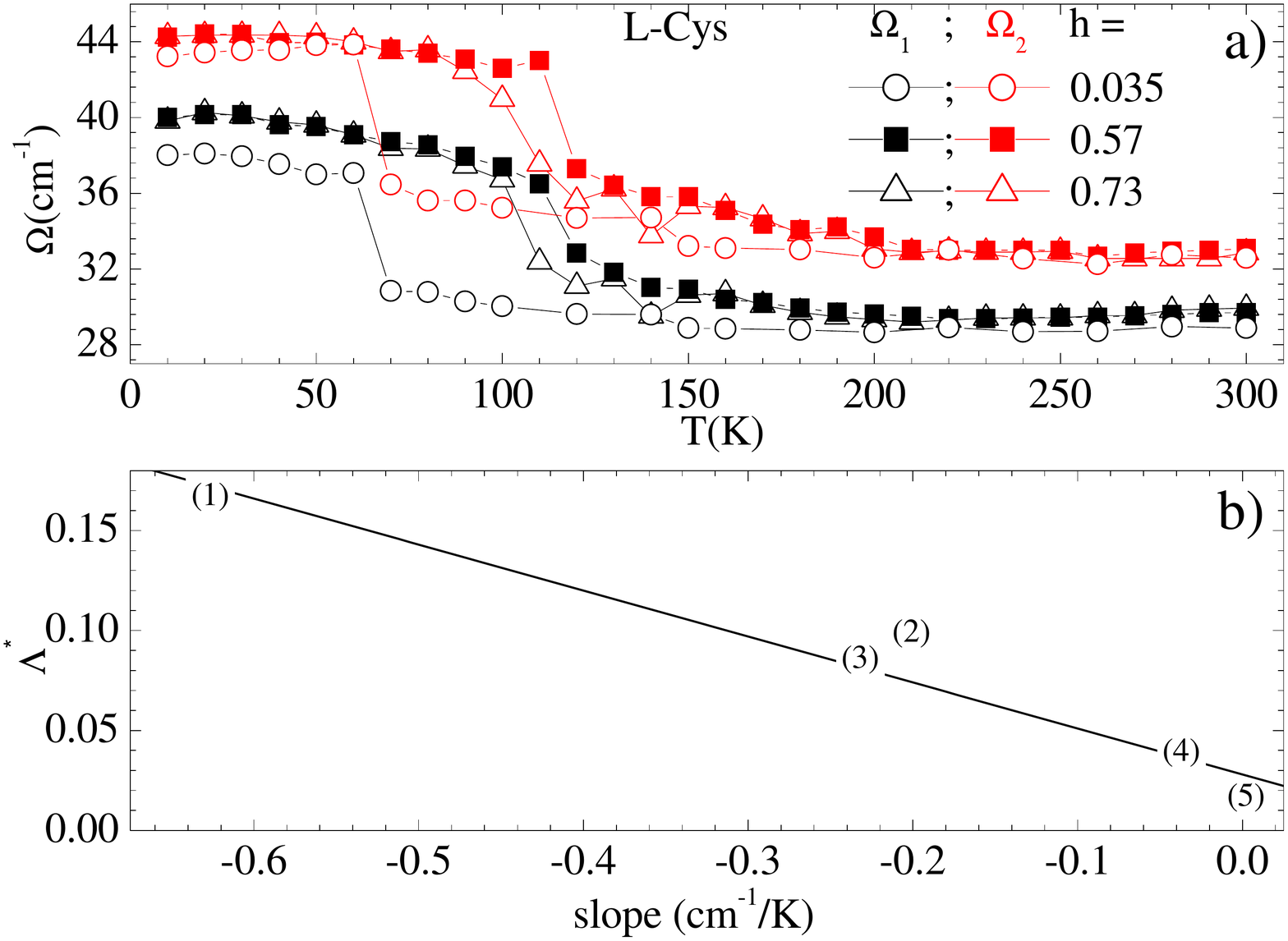}
  \caption{a) Temperature dependence of $\Omega$ parameter for $h = 0.035$ (circles); $0.57$ (squares); and $ 0.73$ (triangles)  L-Cys samples. Black and red symbols refer to behavior of peaks at $\Omega_{1}$ and $\Omega_{2}$ , respectively.b) Quantumness, $\Lambda^{*}$ as function of $BP$ slope. The straight line is a fit to $\Lambda^{*}=0.028-0.23\times slope$.  (1): L-Cys$0.035$; (2):LCys$:0.57$; (3):L-Cys$:0.73$; (4):Lyzozime (refs. \cite{miyazaki2000low,khodadadi2010broad} ); (5):Silica Molten (refs. \cite{molten2006,silicadebye} ).}
  \label{fig:3}
\end{figure}

Figure \ref{fig:3} presents the  temperature and hydration dependence of the parameter  $\Omega$ .  $\Omega(T)$ (Fig. \ref{fig:3}a) presented an abrupt step-like softening with maximum derivative at $T\sim 64$ K in the almost dry sample for both $\Omega_{1}$ and $\Omega_{2}$. Similar behavior was observed in glass-forming liquids such as ortho-terphenyl (OTP) near its glassy transition temperature $T_{g}$.\cite{steffen1994depolarized} However, Khodadadi \textit{et al.},\cite{khodadadi2010broad}, and Sokolov \textit{et al.} \cite{sokolov1994dynamics}  reported the temperature dependence for the $BP$ frequency of lysozyme hydrated glass as broad transition. It is well established that the abrupt softening is a signature of the glassy transition occurrence. Thus one conclude that for L-Cys$:0.035$ $T_{g}\sim 64$K. The observed softening broadened and shifted to higher temperatures with hydration. For $h=0.57$ and $0.73$  it was observed at $T_{g}=115$ and $T_{g}= 110$ K, respectively.

Since the melting temperature ($T_{m}$) for L-Cys  is $493$ K, the glassy ratio $ T_{g}/T_{m}=0.12$ is smaller than the classical value of $2/3$.\cite{thermoglassy} At higher hydration levels,  $T_{g}/T_{m}=0.20$. Novikov and Sokolov\cite{novikov2013role} shown that quantum effects could lead to a significant decrease in the  $T_{g}/T_{m}$ ratio. These effects will have significant influence when the ratio $T_{g}$ to Debye temperature ($\Theta_{Debye}$)  $T_{g}/\Theta_{D}< 0.5$ where thermal ($u(T)$) and zero-point ($u_{0}$) atomic displacements become comparable.  Since $\Theta_{D}=196$ K for L-Cys \cite{ishikawa2013} it was found $T_{g}/\Theta_{D}=0.30$ for L-Cys$:0.035$ and   $T_{g}/\Theta_{D}=0.51$ for the others samples.  Thus, quantum effects are expected to be important at lower hydration levels the most.  A correction factor of $(1+\Theta_{D}/4T_{g})$ to the classical value was proposed by Novikov and Sokolov (see eq. (5) of ref. 32 ) to account for the quantum effects so that

\begin{equation}\label{tgtm}
(\frac{T_{g}}{T_{m}})_{teo}\approx\frac{2/3}{1+\frac{\Theta_{D}}{4T_{g}}}
\end{equation}

Thus $(T_{g}/T_{m})_{teo}=0.37$ and $0.45$ for LCys almost dry and hydrated samples, respectively. These values are greater than the experimental value of $0.12$. We argue that the origin for this discrepancy relies to the excess of VDOS due to the $BP$ neglected on deriving eq. \ref{tgtm} as pointed out by Novikov and Sokolov. Based on supplementary material of ref. \cite{novikov2013role}, it was estimated that $u(T)$ and $u_{0}$ were comparable at $T/\Theta_{D}\sim0.25$ when taken into account the $BP$ of L-Cys. Hence, the role of the zero-point quantum fluctuations will be dominating in this case.

It is possible to include a phenomenological factor in eq. \ref{tgtm} which could be useful for one comparing theoretical findings and experimental data. It is proposed to re-write eq. \ref{tgtm} as

\begin{equation}\label{tgtm2}
\frac{T_{g}}{T_{m}}=\frac{2/3}{\alpha(1+\frac{\Theta_{D}}{4T_{g}})}
\end{equation}

The $\alpha$ values for L-Cys almost dry and hydrated were $3.01$ and $2.01-2.14$, respectively. For glasses were $BP$ contributions would be negligible $\alpha=1$. It was also observed that both $\Omega_{1}(T)$ and $\Omega_{2}(T)$ become almost constant above $T\sim230$ K. This temperature coincides with the dynamic transition temperature $T_{D}$ for L-Cys.\cite{lima2012anharmonic}.

It has been recognized that quantum fluctuations can promote or inhibit glass formation.\cite{novikov2013role, glassquan2011}.  The dimensionless parameter quantumness,$\Lambda^{*}$, defined as the ratio of the de Broglie thermal wavelength to the particle size, controls the scale of quantum behavior. At glass transition it could be estimated $\Lambda^{*}(T_{g})\thickapprox 0.05 \Theta_{D}/T_{g}$.\cite{novikov2013role} It was estimated the corresponding $\Lambda^{*}$ value for the samples presented in this work and some data available on literature for lysozyme\cite{miyazaki2000low,khodadadi2010broad} and molten silica\cite{molten2006,silicadebye} as well and compared to the slope of the $BP$ frequency $\omega$ at $T_{g}$. (Fig.\ref{fig:3}b) It is noticed a remarkable linear dependence between these two parameters, as could be observed on Fig. \ref{fig:3}b). It could be stated that higher the quantumness of a glass sharper is the glassy transition probed by $BP$. Therefore, the quantum fluctuations in the glassy state could be directly probed by the $BP$ Raman spectra.

\begin{figure}[tbh!]
\centering
  \includegraphics[height=5.5cm]{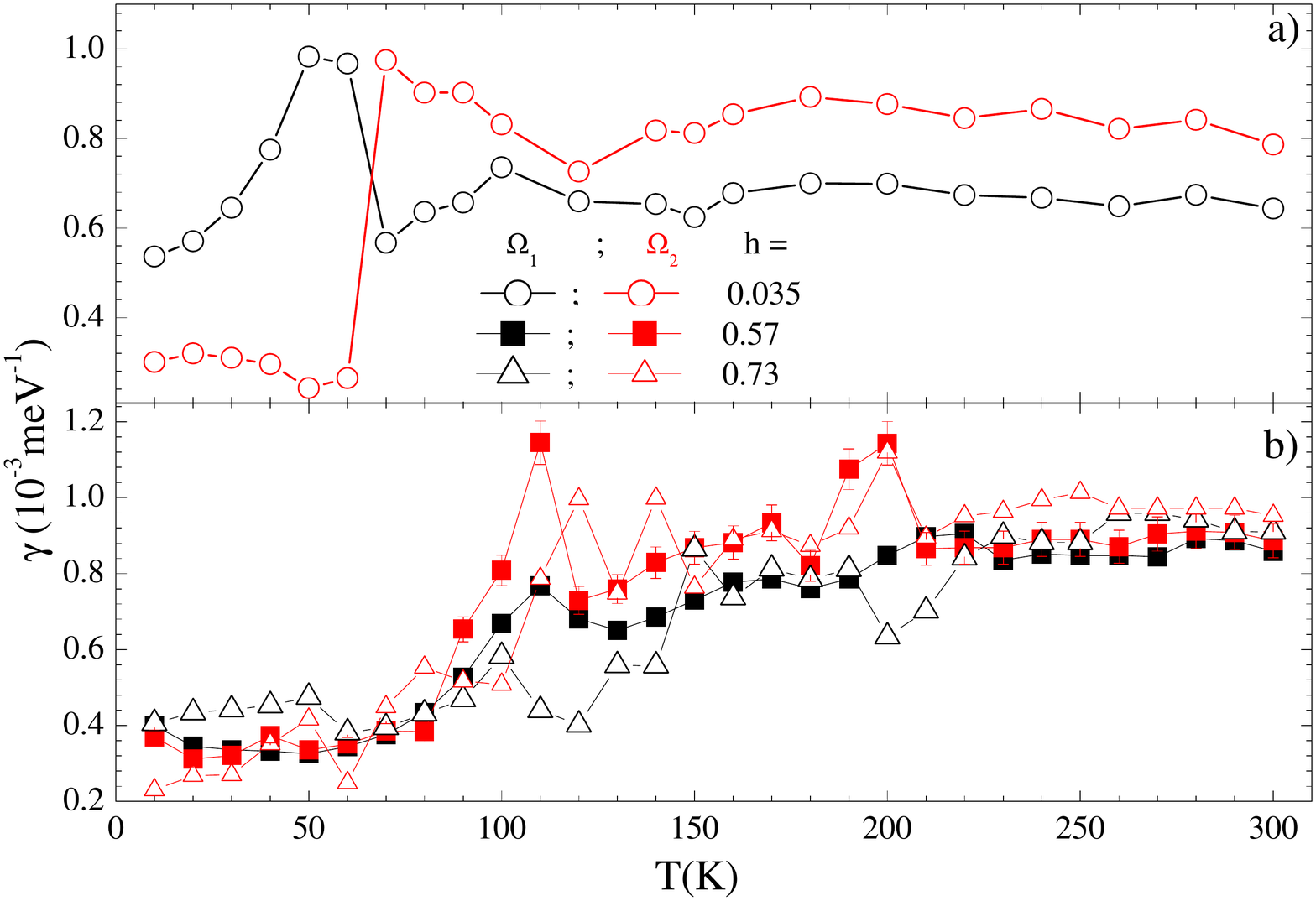}
  \caption{Temperature dependence of $\gamma$ parameter for $h = 0.035$ (circles)(a); $0.57$ (squares)(b); and $ 0.73$ (triangles)(b)  L-Cys samples. Black and red symbols refer to behavior of peaks at $\Omega_{1}$ and $\Omega_{2}$ , respectively.}
  \label{fig:4}
\end{figure}

\begin{figure}[tbh!]
\centering
  \includegraphics[height=4cm]{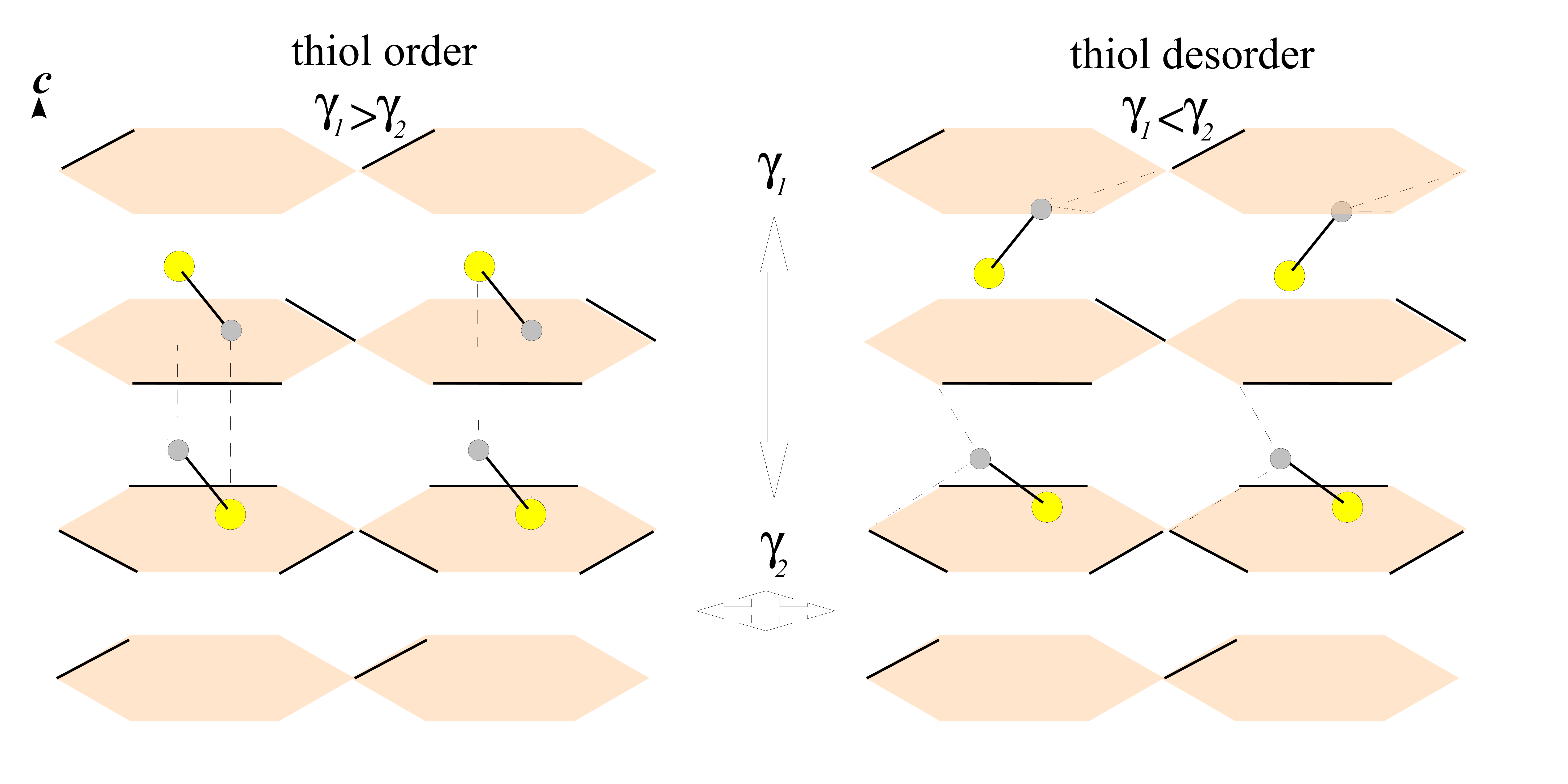}
  \caption{Effect of the thiol group ordering  and disordering  on the relative mobility of perpendicular ( $\gamma_{1}$) and in plane ($\gamma_{2}$) "hexagonal motif".}
  \label{fig:5}
\end{figure}

Figures \ref{fig:4}a) and b) show the temperature behavior of Newtonian friction $\gamma$ parameter for almost dry and hydrated samples, respectively. It was observed that the $h=0.035$ sample (Fig. \ref{fig:4}a) has a very diverse thermal evolution compared to the others. $\gamma_{1}(T)$ and $\gamma_{2}(T)$ 	followed each one distinctive temperature dependence bellow $T_{g}\sim 65$ K in this case. On the other hand, the temperature dependence of $\gamma_{1}$ and $\gamma_{2}$ were almost the same for $h=0.57$ and $0.73$ (Fig. \ref{fig:4}b). The distinctive behavior observed in almost dry sample enable one to associate $\gamma_{1}(T)$ and $\gamma_{2}(T)$ to two different $BP$s each one connected to specific structural features of L-Cys. Below to $T=80$ K and close to $T_{g}$ it was reported\cite{moggach2005cysteine} the ordering of the thiol groups along the $c-$axis.  This ordering connects neighbor hydrogens and sulphur atoms with hydrogen bonding. Each sulphur site is located on the surrounding of a "hexagonal motif" when projected on the $ab-$plane.\cite{moggach2005cysteine} The structure could be viewed as a packing of these motifs along $c-$axis forming tube-like structures connected by hydrogen bonding. From the molecular flexibility point of view one could separate the molecule into two well-defined portions. One related to the possible dislocations along $c-$axis which will be characterized by a friction coefficient $\gamma_{1}$. The other to the in-plane movement, $\gamma_{2}$ (see Fig. \ref{fig:5}). The thiol ordering confines the hydrogen of thiol group to bond along neighbor planes. This enhance the interaction between hexagonal motifs which manifest as increased friction coefficient $\gamma_{1}$. If one directly associates $\gamma_{1}$ to the $\gamma(\Omega_{1})$ whose behavior is displayed on Fig. \ref{fig:4}a) the interpretation of its temperature dependence is direct. Opposite behavior will be expected to the friction coefficient between neighbor tubes ($\gamma_{2}$). The ordering weak the interaction among tubes decreasing $\gamma_{2}$. Again, the association $\gamma_{2}\rightarrow\gamma(\Omega_{2})$ enable one to suitable qualitative interpretation of Fig.\ref{fig:4} b). The directional character of the hydrogen bonding of thiol lost their importance with hydration. The hydrogen bonding due to the water randomly distributed inside the structure will be responsible for homogenously enhance the in-plane and between tubes interactions. Thus, as seen on Fig.\ref{fig:4}b) $\gamma_{1}\sim\gamma_{2}$ for hydrated L-Cys samples. This interpretation could be extrapolated to other glasses or glass-like materials were eventually a two-peaks spectral structure of the $BP$ is observed. In these cases a direct link to the anisotropy of the elastic properties (e.g., $\gamma_{1}\neq\gamma_{2}$) could be invoked to explain the $BP$ spectra. 

\textsc{\textbf{Acknowledgements}} The authors are grateful to the Brazilian agencies FAPESP, CAPES, and CNPq for the financial support and the Multiuser Central Facilities of UFABC for the experimental support.

\bibliography{amino}

\end{document}